\documentclass[12pt]{article}

\usepackage{scicite}

\usepackage{times}
\usepackage{graphicx}
\usepackage{color}
\usepackage{bm}
\newenvironment{sciabstract}{%
\begin{quote} \bf}
{\end{quote}}

\title{Room temperature coherent control of protected qubit in hexagonal boron nitride}

\author
{ Andrew J. Ramsay$^{1}$, Reza Hekmati$^{2}$, Charlie J. Patrickson$^{3}$, Simon Baber$^{3}$, \\ David R. M. Arvidsson-Shukur$^{1}$, Anthony J. Bennett$^{2,4}$ and Isaac J. Luxmoore,$^{3,\ast}$ \\
\normalsize{$^1$ Hitachi Cambridge Laboratory, Hitachi Europe Ltd., Cambridge CB3 0HE, UK}\\
\normalsize{$^{2}$School of Physics and Astronomy, Cardiff University, Queen’s Building, CF24 3AA, UK}\\
\normalsize{$^{3}$College of Engineering, Mathematics and Physical Sciences, University of Exeter, EX4 4QF, UK}\\
\normalsize{$^{4}$School of Engineering, Cardiff University, Queen’s Building, CF24 3AA, UK}\\
\\
\normalsize{$^\ast$To whom correspondence should be addressed; E-mail:  i.j.luxmoore@exeter.ac.uk}
}

\date{}

\begin{document}


\baselineskip24pt


\maketitle


\begin{sciabstract}
  Spin defects in foils of hexagonal boron nitride are an attractive platform for magnetic field imaging, since the probe can be placed in close proximity to the target. However, as a III-V material the electron spin coherence is limited by the nuclear spin environment, with spin echo coherence time of $\bm{\sim100~\mathrm{ns}}$ at room temperature accessible magnetic fields. We use a strong continuous microwave drive with a modulation in order to stabilize a Rabi oscillation, extending the coherence time up to $\bm{\sim4~\mathrm{\mu s}}$, which is close to the 10-$\mathrm{\mu s}$ electron spin lifetime in our sample. We then define a protected qubit basis, and show full control of the protected qubit. The coherence times of a superposition of the protected qubit can be as high as $\bm{0.8~\mathrm{\mu s}}$. This work establishes that boron vacancies in hexagonal boron nitride can have electron spin coherence times that are competitive with typical NV-centers in small nanodiamonds under ambient conditions.
\end{sciabstract}

% In setting up this template for *Science* papers, we've used both
% the \section* command and the \paragraph* command for topical
% divisions.  Which you use will of course depend on the type of paper
% you're writing.  Review Articles tend to have displayed headings, for
% which \section* is more appropriate; Research Articles, when they have
% formal topical divisions at all, tend to signal them with bold text
% that runs into the paragraph, for which \paragraph* is the right
% choice.  Either way, use the asterisk (*) modifier, as shown, to

\section*{Introduction}

%hBN is a wide-bandgap Van der Waals material used as an insulator in two-dimensional electronic devices. In 2016, Tran {\ it et al } (\cite{Tran}) reported that defects in hBN can exhibit narrowband single photon emission at room temperature, with more recent works suggesting that Transform limited emission may even be possible. More recently, optically detected magnetic resonance in ensembles of boron vacancies, and even single possibly carbon-related defects has been reported. This implies that hBN also have a spin resource that could be used for applications such as magnetic field sensors, etc.

%hBN is a III-V material, and the interactions with the nuclear spin environment are expected to lead to strong electron spin decoherence, making defects in hBN a less than obvious choice of material for optical spintronic applications.

%So far, most work has focused on the boron vacancy since this is a comm been clearly identified

%ODMR applications in sensors and etc
%Use group IV isotopically purified extraordinary coherence times
%, but in dirty material not so good.

%III-V is not a natural choice. However, interest in hBN because comeptitive with nanodiamonds

Hexagonal boron nitride (hBN) is a wide-bandgap Van der Waals material used as an insulator in two-dimensional electronic devices \cite{Dean_NatNano2010}. Recently, there has been a growing interest in optically detected magnetic resonance (ODMR) experiments on spin defects. There are several reasons why. Firstly, some color-centers in hBN exhibit attractive quantum emission properties such as high brightness at visible wavelengths, well-matched to silicon detectors \cite{Tran_NNano2016}, and suggestions of transform-limited transitions with a high fraction of emission occurring through the zero-phonon line \cite{Hoese_SciAdv2020,Dietrich_prb2020}. ODMR provides a useful tool for identifying defects, that has been instrumental in identifying the boron vacancy in hBN \cite{Gottscholl_NMater2020,Haykal_ArXiv}. Secondly, if a defect could be identified that combines the excellent quantum emitter properties with a spin memory this could provide a competitive system to build a spin-photon interface \cite{Atature_Review}. Thirdly, as a two-dimensional material it may be possible to make magnetic-field sensing foils that allow the spin-defect to be placed in close-proximity to the target of interest \cite{Gottscholl_NComm2021}. In the leading platform, nanodiamonds, the spin coherence properties are adversely affected by the surface states, typically limiting spin echo coherence times to a few $\mathrm{\mu s}$ for diameters of a few nm \cite{Doherty_PhysRep2013}. In hBN, there are a few recent studies on single bright defects \cite{Stern_Ncomm2022,Chejanovsky_Namter2021,Guo_ArXiv2022}, which are possibly carbon related. So far, most of the work in hBN has focused on ensembles of boron vacancies \cite{Gottscholl_NMater2020,Gottscholl_SciAdv2021,Gottscholl_NComm2021,Baber_NL2022,Reimers_PRB2020,Kianinia_ACSPhoton2020}, because although they suffer low brightness, they are easy to generate \cite{Kianinia_ACSPhoton2020}, their internal energy levels have been theorized \cite{Reimers_PRB2020,Abdi_ACSPhoton2018,Ivady_NPJCM2020} and allow facile spin-pumping by a green laser.  

In any III-V material the nuclear spin environment is expected to limit the electron spin coherence times. So far, in hBN there have been claims of spin echo times in the $\mu s$-regime, at room temperature and moderate magnetic-field \cite{Gottscholl_SciAdv2021,Gao_NL2021}, which is competitive with NV-centers in small nano-diamonds \cite{Doherty_PhysRep2013}.  However, this has been challenged by experiments that show sub-100 ns spin echo times in isotopically purified material, which are further supported by calculations of the decoherence expected from the electron-nuclear interactions \cite{Haykal_ArXiv}.

Here, we present a study of the spin properties of ensembles of $V_B^-$ in hBN. We confirm the findings of Haykal {\it et al} \cite{Haykal_ArXiv} that the majority of the spin echo coherence is lost in $T_{echo}<100~\mathrm{ns}$ at room temperature and milli-Tesla magnetic fields. However, we propose and demonstrate a solution. We show that by using a strong continuous microwave field, a method often referred to as continuous concatenated driving (CCD), we can define a protected qubit basis with favorable coherence properties, allowing the Rabi oscillation damping time to be extended up to $4~\mathrm{\mu s}$. This method has previously been used in NV-centers in diamond \cite{Wang_NJP2020,Cai_NJP2012,Rohr_PRL2014,Stark_Ncomm2017,Cao_PRApp2020,Wang_PRA2021,Farfurnik_PRA2017}, but here we show that it works extremely well in a III-V material with hostile nuclear environment. Furthermore, we define a protected qubit basis in terms of an electron spin that rotates in-phase or out-of-phase with the expected Rabi oscillation. We then demonstrate full control of this protected qubit to show coherence times, $T_{pRabi}$, of up to $0.8~\mathrm{\mu s}$. Using this protection scheme, we therefore show that spin defects in hBN can have useful microsecond-scale coherence times at room temperature and milli-Tesla magnetic field.

Our device consists of flakes of hBN placed on top of a co-planar waveguide (CPW) fabricated on a sapphire substrate \cite{Baber_NL2022}. Following the recipe of ref. \cite{Kianinia_ACSPhoton2020}, boron vacancies are generated by carbon-ion irradiation at 10 keV and a dose of $1\times 10^{14}~\mathrm{cm^{-2}}$. A d.c. magnetic field of 20 mT is applied along the c-axis of the hBN flakes, and an a.c. field is applied in-plane through the CPW. The coupling is sufficient to achieve Rabi frequencies in excess of 100 MHz. The sample is located beneath a microscope at room temperature and in air. Photoluminescence (PL) is excited using a 532 nm laser, modulated using an acousto-optic modulator, and detected with a Si-APD module.

Only the negatively charged boron vacancy ($V_B^-$) is expected to be optically active in the range of our detection system \cite{Comtet_NL2019}. A sketch of the optical pumping cycle for optically detected magnetic resonance (ODMR) is shown in Fig. \ref{fig:fig1}A. The $V_B^-$ has two unpaired electrons in an S=1 triplet ground state. Excitation with a 532 nm laser preferentially optically pumps the $V_B^-$ into the $m_s=0$ state, whose PL is slightly brighter than the $m_s=\pm 1$ states, allowing ODMR detection of the electron spin resonance. The PL is broadband and centered at $\sim850$ nm, see Fig. \ref{fig:fig1}B.

The Hamiltonian of the crystal ground-state can be expressed as $H=H_e+H_n+H_{en}$. The electron Hamiltonian, 
\begin{equation}
H_e=DS_z^2+E(S_x^2-S_y^2) +\gamma_e B_zS_z,
\end{equation}
is composed of a zero-field splitting with $D=3.47~\mathrm{GHz}$, and $E=150~\mathrm{MHz}$ \cite{Baber_NL2022}, and an electron Zeeman term with gyromagnetic ratio $\gamma_e=28~\mathrm{MHz/mT}$. $S_j$ are the S=1 electron spin operators.
 D is consistent with previous works \cite{Gottscholl_NMater2020}. $E$ is relatively large, perhaps due to high level of strain caused by use of carbon-irradiation to generate the defects. The nuclear spin Hamiltonian is
 \begin{equation}
 H_n=\sum_{k}\gamma_n^kB_zI_z^k +\mathbf{I^kQ^kI^k}.
 \end{equation}
   $I^k$ is the nuclear spin of nuclei $k$. The first term is a nuclear Zeeman term, and $\mathbf{Q^{k}}$ is a quadrupolar tensor. $H_{en}$ is the electron-nuclear hyperfine interaction.
  \begin{equation}
  H_{en}=\sum_k \mathbf{SA^kI^k}
  \end{equation}
   It is dominated by the three nearest neighbor nitrogen atoms with $A^{nn}=47~\mathrm{MHz}$, and $I(^{14}N)=1$ \cite{Gottscholl_NMater2020}. The next nearest neighbor interactions have been calculated to be up to 6.8 MHz in ref. \cite{Ivady_NPJCM2020}. In our CW-ODMR measurements, see Fig. \ref{fig:fig1}C, the hyperfine structure is not resolved, but the FWHM is similar to $3A$, corresponding to three nearest neighbors with $I=1$.

\section{Unprotected spin}

\begin{figure}%
\includegraphics[width=1\columnwidth]{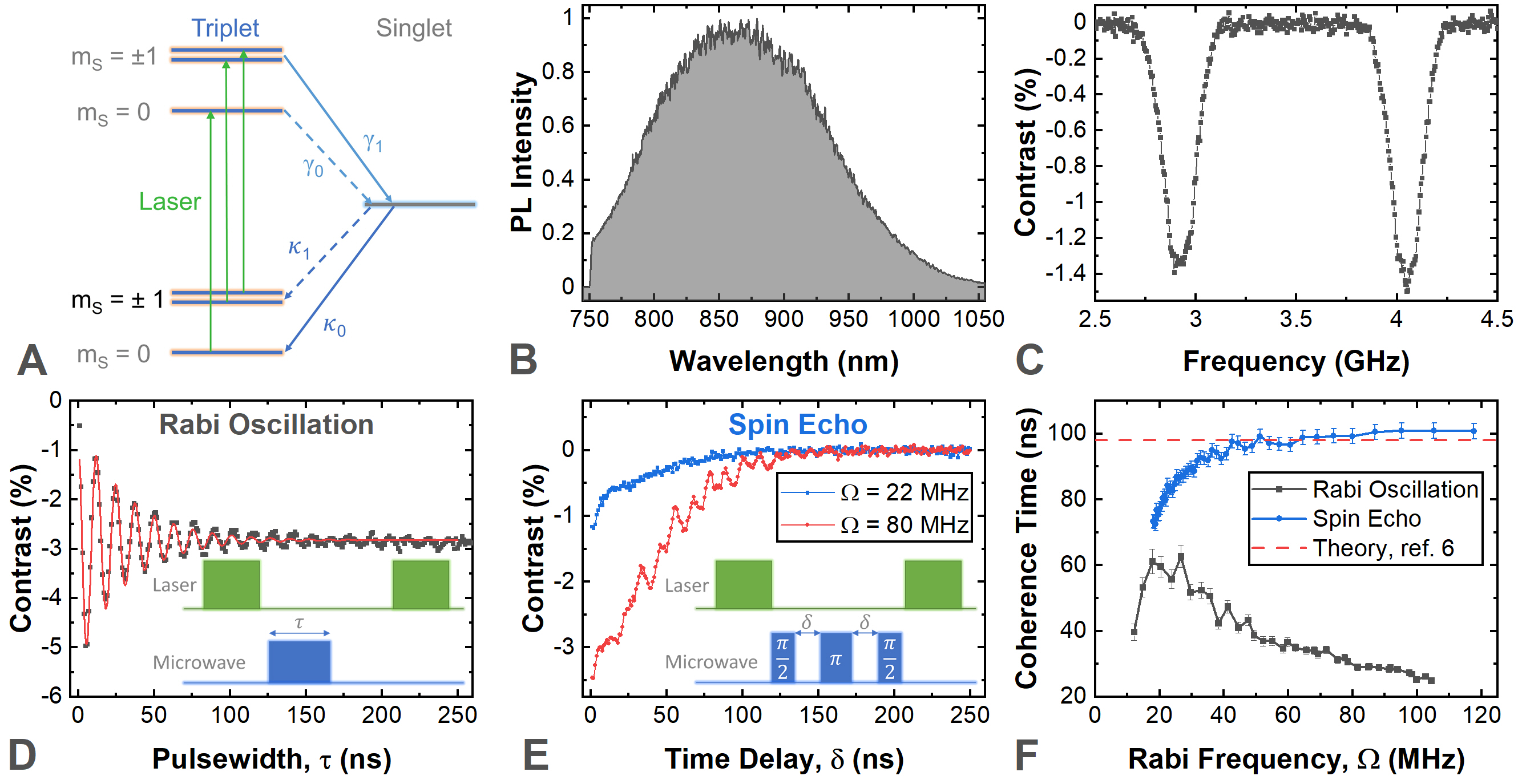}
\caption{ Coherent control of unprotected boron vacancy spins. (\textbf{A}) Energy levels involved in ODMR of the boron vacancy in hBN. Optical pumping combined with spin dependent inter-system crossing, initializes the system in $m_s=0$ triplet ground state. This is slightly brighter than the $m_s=\pm 1$ states, allowing detection of the spin state. (\textbf{B}) Photoluminescence under 532 nm excitation. (\textbf{C})  Continuous wave ODMR spectrum, the FWHM of 150 MHz is mostly due to unresolved hyperfine structure. (\textbf{D}) Rabi oscillation of the $m_S = 0$ to $m_S = +1$ ground state transition (\textbf{E}) Comparison of spin echo at low (22 MHz, black) and high (80 MHz, red) Rabi frequencies. At high Rabi frequency, the contrast is stronger since higher proportion of spins are controlled. (\textbf{F}) Comparison of coherence times extracted from Rabi oscillation and spin echo measurements, as a function of Rabi frequency. For Rabi oscillations the coherence time decreases at high Rabi frequency due to fluctuations in drive strength \cite{Valla_PhysRevAppl2021}. For spin echo, the coherence time saturates at high Rabi frequencies, where the pulse is shorter relative to the spin precession time. The saturation value is close to that calculated by Haykal {\it et al} (dashed-line) for hBN with natural isotope distribution, \cite{Haykal_ArXiv}. The Rabi oscillations are fitted to $f(\tau)=a+be^{-\tau/T_{Rabi}}\cos{(\Omega \tau)}$, and spin echo to $g(\delta)=-a.e^{-2\delta/T_{echo}}$. 
}
\label{fig:fig1}%
\end{figure}

To start our investigation of the spin properties of $V_B^-$ we measure a Rabi oscillation, see Fig. \ref{fig:fig1}D. The Rabi oscillation is sensitive to all inhomogeneities and noise sources, and has a coherence time of $T_{Rabi}<60~\mathrm{ns}$, which depends on the microwave power (Fig. \ref{fig:fig1}F). At high microwave power, the dephasing rate is proportional to the Rabi frequency possibly due to fluctuations in the power of the microwave source. At low  microwave power, the Rabi damping is dominated by fluctuations in the detuning \cite{Valla_PhysRevAppl2021}.

To evaluate the intrinsic coherence times, we perform spin-echo measurements, see Fig. \ref{fig:fig1}E. In principle, the measurement is insensitive to low frequency variations in the detuning, and errors in the pulse-area. However, at a moderate magnetic field of B = 20 mT, we find a $T_{echo}<100~\mathrm{ns}$, that saturates at high Rabi frequencies with fast pulses. We note that the $T_{echo}$ is comparable to the $T_{Rabi}$ times. These findings are similar to those of Haykal {\it et al} \cite{Haykal_ArXiv}, and in the high microwave-power regime our $T_{echo}=100~\mathrm{ns}$ is a close match to their calculations for a natural 20:80 mix of $^{10}B$ and $^{11}B$ isotopes. We note that data reported by Liu {\it et al} \cite{Liu_ArXiv2021} is compatible with $T_{echo}$ of $<$100 ns under similar conditions. In refs. \cite{Gottscholl_SciAdv2021,Gao_NL2021} a longer $\mu s$-scale $T_{echo}$ was reported. However, these $T_{echo}$ were extracted from low contrast data for times larger than 200 ns. We conclude that the majority of the electron spin coherence is lost in the first 100 ns, and hypothesize that a small fraction of the coherence may remain on a longer time-scale. At few Tesla magnetic fields, $T_{echo}=15~\mathrm{\mu s}$, similar to the $T_1$ time, has been reported \cite{Murzakhanov_NanoLett2022}. However, for practical purposes the use of superconducting magnets renders the room temperature aspect of the system academic.

\section{Stabilized Rabi oscillation}

%Compared to other III-V systems, such as GaAs quantum dots \cite{Greilich_sci03,GaAselec}, a $T_{2,SE}=100~\mathrm{ns}$ at moderate magnetic field, and room temperature, offers  advantages for practical sensing applications \textcolor{red}{check}. However, 
An important question for quantum devices based on hBN spin defects is whether one can create a protected qubit that is isolated from the nuclear spins, and thereby access a longer electron spin coherence. There exists an extensive literature on how to mitigate the impact of nuclear spins in III-V quantum dots \cite{Chekhovich_NMater2013,Gangloff_sci2019}, and for defects in group-IV materials \cite{Miao_Sci2020,Wang_PRA2021,Wang_NJP2020}. For the boron vacancy, the electron-nuclear interaction is dominated by the three nearest nitrogen atoms and a strong hyperfine interaction of 47 MHz. This few nuclei situation, contrasts with the case of a GaAs quantum dot, where a few thousand nuclei provide a bath of nuclear spins. It also contrasts with the case of group IV-materials, where the majority of nuclei have no nuclear spin, and electron spin dephasing due to hyperfine coupling is relatively weak.

To this end, we trial the use of a strong microwave field to create a protected spin-qubit subspace, and use concatenated continuous driving (CCD) methods to coherently control the protected qubit \cite{Wang_PRA2021,Wang_NJP2020}. This involves applying a continuous ac magnetic field along the x-direction to give a control Hamiltonian of the form:
\begin{eqnarray}
H_c(t)=[\Omega\cos{(\omega t+\phi)}+2\epsilon_m\sin{(\omega t+\phi)}\sin{(\omega_m t-\theta_m)}]S_x
\end{eqnarray}
The first term is the usual Rabi drive. The second term adds an amplitude modulated field in quadrature with the Rabi drive, and when $\omega_m=\Omega$ this acts to stabilize the Rabi oscillation. This can be understood by considering only the $m_s=0, m_s=+1$ electron spin states near resonance with the Rabi drive, and switching to the first rotating frame of the Rabi drive:  $H_c^{'}=e^{i\omega t\sigma_z/2}H_ce^{-i\omega t\sigma_z/2}$. In the case $\phi=0$,
\begin{equation}
H_c^{'}=\frac{1}{2}[\Omega\sigma_x^{'}-2\epsilon_m\sin{(\omega_m t -\theta_m)}\sigma_y^{'}] \label{eq:epsfield}
\end{equation}
where $\sigma_{\alpha}^{'}$ are the Pauli spin-1/2 matrices, and the superscript identifies the frame. The counter-rotating term has been neglected. The first term gives rise to a Rabi oscillation in the yz-plane. The second term applies a corrective rotation about the y-axis. This is maximum when the Bloch-vector is scheduled to point along y-direction, and if there is an error it will rotate the spin closer to the target value. This results in a kind of spin-locking effect that can sustain the Rabi oscillation.

Further insight can be gained by making a second rotating frame approximation to rotate into the dressed states basis, $H_c^{''}=e^{i\Omega t\sigma^{'}_x/2}H_c^{'}e^{-i\Omega t \sigma^{'}_x/2}-\Omega\sigma_x^{'}/2$.
\begin{eqnarray}
H_c^{''}=\frac{\epsilon_m}{2}[\sin{(\theta_m)}[-\sin{(\phi)}\sigma_x^{''}+\cos{(\phi)}\sigma_y^{''}]+\cos{(\theta_m)}\sigma_z^{''}]
\label{eq:secondfield}
\end{eqnarray}
In the protected frame, the control field is equivalent to a d.c. magnetic field of magnitude $\epsilon_m$ that is oriented using the phases of the drive $\theta_m$ and $\phi$. In the lab frame, the protected states $\vert 0^{''}\rangle$, $\vert 1^{''}\rangle$ correspond to an in-phase/out-of-phase Rabi oscillation, respectively. If the drive is set such that $\theta_m=0$, the Rabi oscillation is the state $\vert 0^{''}\rangle$, and is protected by an effective stabilizing magnetic field of magnitude $\epsilon_m$ that is aligned anti-parallel to the target Bloch-vector, this induces an energy gap $\epsilon_m$ over the full Bloch-vector trajectory of the Rabi oscillation.

\begin{figure}%
\includegraphics[width=1\columnwidth]{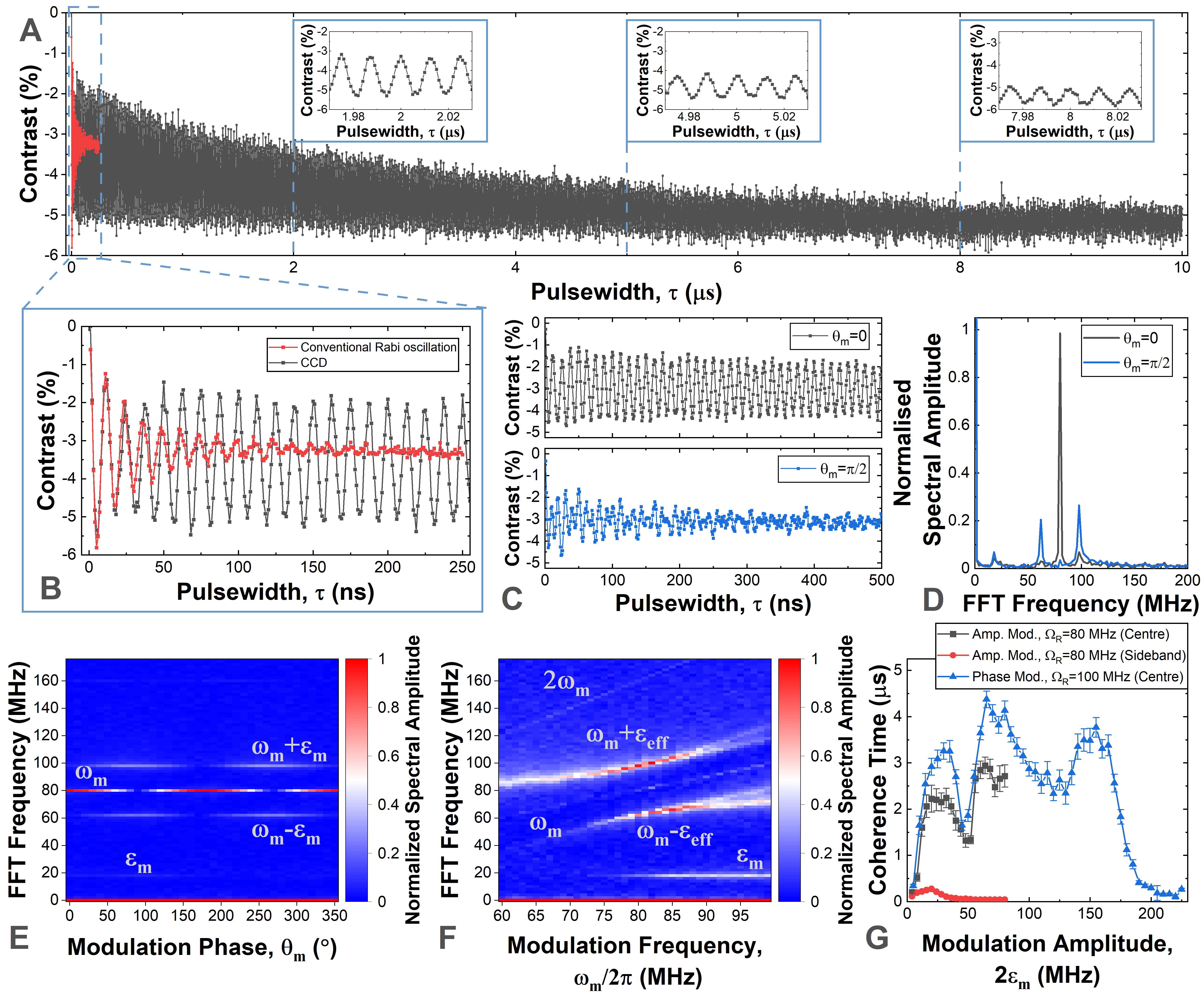}
\caption{Stabilization of Rabi oscillation. (\textbf{A, B}) Comparison of unprotected Rabi oscillation (red) and amplitude modulated CCD scheme with $\epsilon_m=12~\mathrm{MHz}$ (gray), showing stabilization of Rabi oscillation on $\mathrm{\mu s}$-timescale. (\textbf{C}) Phase control of amplitude modulated CCD scheme. If $\theta_m=0$ the stabilization works, and the spin oscillates at single Rabi frequency. If $\theta_m=\pi/2$, the $\epsilon_m$-term modulates the amplitude of the Rabi frequency,  resulting in a strongly damped oscillation at the two sideband frequencies $\omega_m\pm\epsilon_m$. (\textbf{D}) Fourier transform spectrum of (\textbf{C}). (\textbf{E}) Frequency spectrum of amplitude modulated CCD Vs modulation phase $\theta_m$, when $\omega_m=\Omega=80~\mathrm{MHz}$ and $\epsilon_m=20~\mathrm{MHz}$. $\theta_m$ tunes between single band at $\omega_m$ and two sidebands at $\omega_m\pm \epsilon_m$. (\textbf{F}) Frequency spectrum of amplitude modulated CCD vs modulation amplitude $\epsilon_m$, with $\theta_m=\pi/2$ and $\epsilon_m=20~\mathrm{MHz}$. The main feature is a Mollow-triplet like structure with components $\omega_m,\omega_m\pm \epsilon_m$. (\textbf{G}) Comparison of coherence times of center-band and sidebands as a function of amplitude modulation $\epsilon_m$.}

\label{fig:fig2}%
\end{figure}

Fig. \ref{fig:fig2}A presents a comparison of unprotected Rabi oscillation (red), and a protected-qubit Rabi oscillation using a CCD drive with $\theta_m=0, \phi=0$ (black), with a close up shown in Fig. \ref{fig:fig2}B. The former is exponentially damped with $T_{Rabi}=31~\mathrm{ns}$. For the CCD scheme, after the first cycle, the amplitude of the Rabi oscillation has stabilized. The stabilized Rabi oscillation is also exponentially damped, but with an extended time of $T_{CCD}=2.2~\mathrm{\mu s}$.

In the Fourier domain, the stabilized Rabi oscillation narrows the $\Omega$ frequency component of the signal, and gives rise to two weak sidebands at $\Omega\pm \epsilon_m$. Fig. \ref{fig:fig2}C compares the stabilized Rabi oscillation in the case with $\theta_m=0,\pi/2$. For $\theta_m=\pi/2$, the stabilization $\epsilon_m$-term adds in phase with the Rabi-term resulting in an amplitude modulated Rabi drive, and an unprotected Rabi oscillation with strong damping. In the Fourier domain, the phase $\theta_m$ controls the magnitude of the center band, and the sidebands, see Fig. \ref{fig:fig2}D. Fig. \ref{fig:fig2}E and Fig. \ref{fig:fig2}F show maps of the modulation phase and frequency ($\omega_m$) dependence of the signal in the frequency domain with $\theta_m=\pi/2$, respectively. The main feature is a Mollow-triplet \cite{Wang_PRA2021} with center frequency $\omega_m$, and sidebands $\omega_m\pm \sqrt{(\omega_m-\Omega)^2+\epsilon_m^2}$. A feature at $\epsilon_m$ and a counter-rotating term at $2\omega_m$ can also be seen. At high $\omega_m$,  additional side bands at $\omega_m\pm \sqrt{(\omega_m-\Omega_{m_I})^2+\epsilon_m^2}$, $\Omega_{m_I}^2=\Omega^2+(m_IA^{nn})^2$, are observed, and attributed to transitions where the nearest neighbor nuclei are in total nuclear spin state $m_I=\pm 1$.

To characterize the lifetime of the stabilized Rabi oscillation, we plot the coherence time of the center and sidebands as a function of stabilization field strength $2\epsilon_m$ (Fig. \ref{fig:fig2}G). For an amplitude modulation with $\theta_m=0$, when $2\epsilon_m>10~\mathrm{MHz}$ the energy-gap of the protected qubit exceeds the next-nearest neighbor electron-nuclear hyperfine coupling strengths of approximately 6.8 MHz \cite{Ivady_NPJCM2020}, and the center-band coherence time increases to better than a microsecond, as the electron spin is isolated from the majority of nuclear spins. At $2\epsilon_m\approx A=47~\mathrm{MHz}$, there is a dip when the $2\epsilon_m$ sideband-splitting matches the nearest-neighbour hyperfine interaction, before rising to a maximum value of $T_{CCD}\approx 2~\mathrm{\mu s}$. The sidebands are far broader, and the optimum coherence time of the sidebands occurs at $2\epsilon_m\approx 20~\mathrm{MHz}$ with $T_{CCD\pm}=0.27~\mathrm{\mu s}$. We interpret the sidebands as providing the stabilization of the Rabi oscillation, and need a faster response to stabilize the electron spin. Using amplitude modulation, $2\epsilon_m$ is limited by the available power. A similar control can be achieved using phase-modulation, where the control Hamiltonian is: $H_c=\Omega\cos{(\omega t +\phi -\frac{2\epsilon_m}{\Omega}\sin{(\omega_m t-\theta_m)})}S_x$. This has the advantage of not requiring so much power, allowing us to extend the range of $2\epsilon_m$. Qualitatively, the coherence time dependence for phase modulation imitates the amplitude modulation case. However, an additional dip in coherence time is observed at about $100-140~\mathrm{MHz}$, which corresponds to $2\epsilon_m\approx 2\Omega-A$. When $\epsilon_m>\Omega$ the stabilization stops working. Overall, the phase modulation scheme works better, possibly because the frequency stability of the function generator is better than the power stability, and reaches a maximum of $T_{CCD}=4.4~\mathrm{\mu s}$, a greater than 150-fold improvement on $T_{Rabi}\approx25~\mathrm{ns}$ of the conventional Rabi oscillation.

\section{Control of protected electron spin}

\begin{figure}%
\includegraphics[width=1\columnwidth]{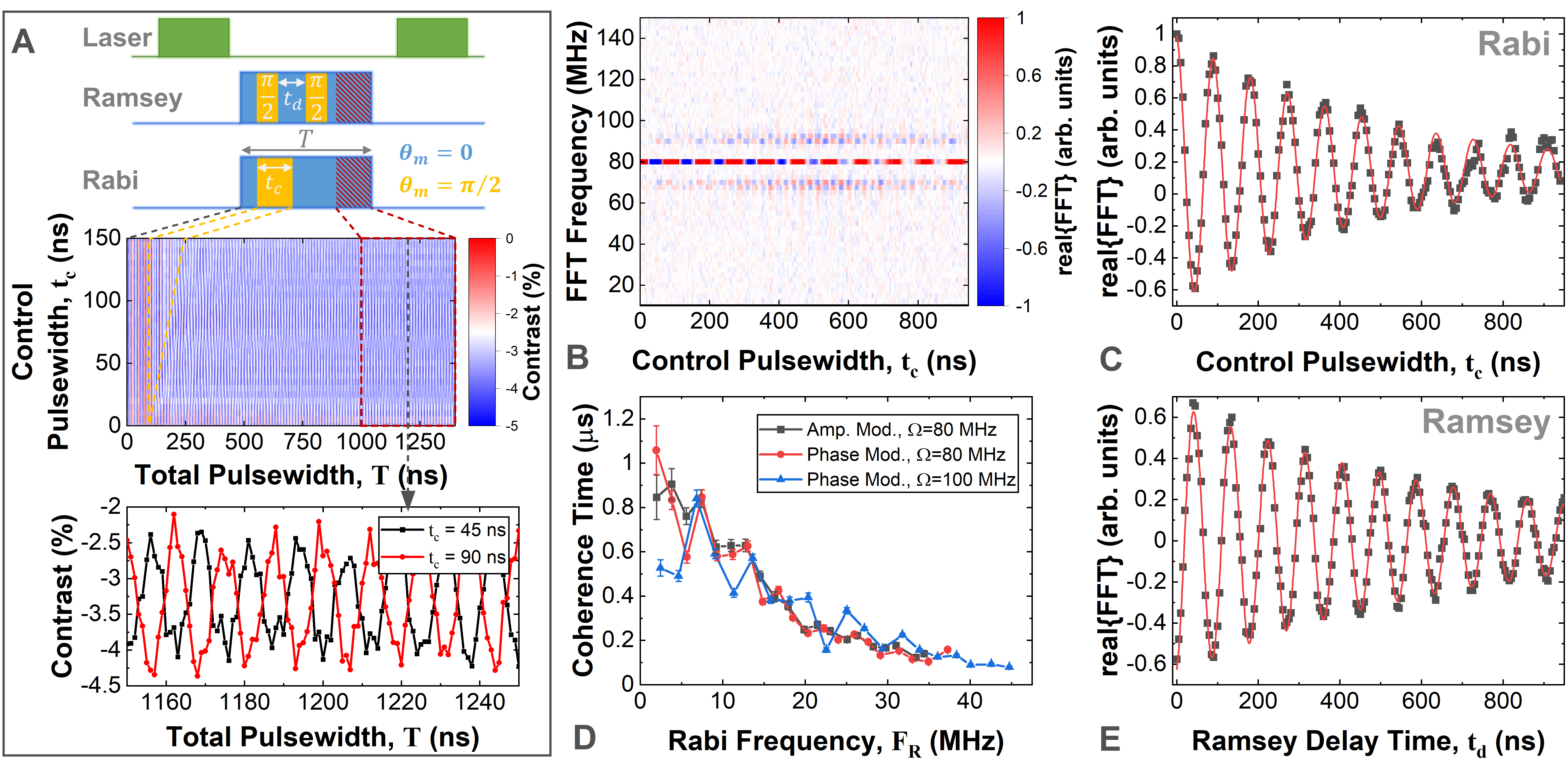}
\caption{ Coherent control of protected qubit. (\textbf{A}) (Upper panel) To perform a Rabi oscillation of the protected qubit the amplitude modulated CCD pulse is applied for a total time $T$. In the dressed states basis, this results in an effective field of magnitude $\epsilon_m/2$, aligned along polar angles ($\theta_m,\phi_m$) as given by Eq. \ref{eq:secondfield}. For the first 94 ns, the protected qubit is prepared by applying $\theta_m=0$, initiating a stabilized Rabi oscillation. The control field is tilted by setting $\theta_m=\pi/2$ for a time $t_c$, and then returned to $\theta_m=0$ to realign the control field along the qubit axis. The qubit state is read out by measuring the phase of the stabilized Rabi oscillation during the time window $t=1000-1400~\mathrm{ns}$. (Middle panel) Measured contrast as a function of total pulsewidth, $T$ and control pulsewidth, $t_c$. (Lower panel) Examples of stabilized Rabi oscillations, recorded during the readout window for $\pi$ ($t_c=45~\mathrm{ns}$) and $2\pi$ ($t_c=90~\mathrm{ns}$) control pulses. (\textbf{B}) Real part of FFT of the readout window in A. (\textbf{C}) The center-band oscillates at protected Protected Rabi frequency $F_R\approx \epsilon_m$. (\textbf{D}) Coherence time of the protected qubit for different pulse schemes and control field amplitudes. The coherence times are best at low protected Rabi frequency, with maximum times of about 800-1000 ns observed. (\textbf{E}) Ramsey interference of protected qubit, using similar measurement protocol shown in (\textbf{A}). }
\label{fig:fig3}%
\end{figure}

So far, we have shown that at room temperature and milli-Tesla magnetic field a freely evolving electrons spin in ensembles of boron vacancies decohere on a timescale of $T_{echo}\approx 100~\mathrm{ns}$ due to a strong electron-nuclear hyperfine interaction. However, by using a CCD scheme it is possible to stabilize the Rabi oscillation of the electron spin, increasing the damping time of the Rabi oscillation up to several $\mathrm{\mu s}$. Regarding applications in metrology or as a spin qubit, the question is, do we simply lock the spin to a Rabi oscillation, or can we define a fully controllable protected qubit, and if so, how coherent is that qubit?

To address this question, we define a protected qubit basis in the second rotating frame. To measure in detail a Rabi oscillation of the protected qubit, we use the measurement scheme shown in Fig. \ref{fig:fig3}A. At time zero, the control field is turned on and set to `idle' with $\theta_m,\phi=0$, and the effective control field $H_c=\frac{\epsilon_m}{2}\sigma_z^{(2)}$ points along the z-axis, see eq. \ref{eq:secondfield}, and the system is initialized in state $\vert 0^{''}\rangle=\cos{(\Omega t/2)}\vert 0^{'}\rangle-i\sin{(\Omega t/2)}\vert 1^{'}\rangle$, which is a stabilized Rabi oscillation starting at time zero. The logical state $\vert 1^{''}\rangle=i\sin{(\Omega t/2)}\vert 0^{'}\rangle+\cos{(\Omega t/2)}\vert 1^{'}\rangle$, and is observed as a Rabi oscillation that is out-of-phase at time zero. If the electron spin is in a superposition state, this will show up as modulation sidebands with frequency component $\Omega\pm \epsilon_m$. To effect a Rabi oscillation, the control field is tilted into $y^{(2)}$-direction by setting $\theta_m=\pi/2$ for time $94~\mathrm{ns}<t<t_c+94~\mathrm{ns}$, and then switched back to `idle'. At the end of the MW-pulse, a laser pulse is switched on for read-out in the lab-frame. We measure the PL contrast as a function of the total MW-pulse width $T$ and the control pulse length $t_c$, see the color map of Fig. \ref{fig:fig3}A. For total time, $T<94~\mathrm{ns}$ a Rabi oscillation is observed indicating the system is in state $\vert 0^{''}\rangle$. For total time, $94~\mathrm{ns}<T<t_c+94~\mathrm{ns}$ the system evolves under the influence of the control pulse, and afterwards the final state is expressed in the Fourier components of the signal. To analyze this we consider the data-set $1000~\mathrm{ns}<T<1400~\mathrm{ns}$, and make a Fourier transform with the phase defined as zero at time $T=0$, see Fig. \ref{fig:fig3}B. The center band oscillates with the control pulse length at a frequency $\epsilon_m$, the sidebands oscillate in quadrature with the center peak, and show up when the protected spin points along the $y^{''}$-direction. As shown in Fig. \ref{fig:fig3}C, the center peak decays with a time $T_{pRabi}=487~\mathrm{ns}$.

To show rotation of the protected spin about a second axis, a Ramsey interference experiment is performed, using the pulse sequence shown in Fig. \ref{fig:fig3}A. The MW-pulse is applied for a total time  $1<T~<1.4~\mathrm{\mu s}$. Again, to initialize the protected spin in state $\vert 0^{(2)}\rangle $ the MW-field is applied in `idle' state for a short time. To apply a $\pi/2$-pulse, $\theta_m$ is switched to $\theta_m=\pi/2$ for a time $\tau_{90}=\pi/2\epsilon_m$, and then switched back to the $\theta_m=0$ idle state. A pair of $\pi/2$-pulses with a time-delay $t_{d}$ are applied, and the final state is deduced by analyzing the signal as the total time $T$ and the time-delay $t_{d}$ are varied, see Fig. \ref{fig:fig3}E. A Ramsey interference showing rotation about $z^{''}$ axis with a frequency of $\epsilon_m$ is observed, with a $T_{pRam}=706~\mathrm{ns}$, demonstrating full control of the protected spin.

To evaluate the potential coherence time of the protected spin, we measure the damping time of the protected Rabi oscillation as a function of the modulation $2\epsilon_m$, see Fig. \ref{fig:fig3}D. The best coherence time $T_{pRabi}=0.8~\mathrm{\mu s}$ is observed at low $\epsilon_m$, which is approximately 8 times better than the unprotected spin echo time.

To conclude, the spin echo coherence time of ensembles of boron vacancies in room temperature hexagonal boron nitride is limited by electron-nuclear interactions to under $100~\mathrm{ns}$ at sub 100 mT B-fields. To overcome this issue, we trial a CCD scheme to stabilize the Rabi oscillation, extending the Rabi damping time up to 4~$\mathrm{\mu s}$, which is close to  $T_1=10~\mathrm{\mu s}$. This time is similar to the spin echo coherence times of InAs/GaAs quantum dots at helium temperatures \cite{Press_nphoton,Greilich_sci}, and NV-centers in 10-35 nm diameter nanodiamonds at room temperature \cite{Knowles_Nmater}.  Here we define a protected qubit basis, show two-axis control, and show that the arbitrary superpositions of the protected qubit can survive for up to 800 ns. The protection scheme uses a high microwave power, but should be insensitive to the details of the nuclear bath, and applicable to other spin defects in hBN or other III-V materials. Future work to understand how this coherence resource can be used for applications in sensing and spin-photon interfaces is required.

\section{Materials and Methods}
\subsection{Sample}
The sample consists of a chromium/gold (20/100 nm thick) coplanar waveguide (CPW), with a $10 ~\mathrm{\mu m}$ wide central conductor, on a sapphire substrate. An hBN flake, approximately 100 nm thick, is placed on top of the CPW using the PDMS transfer method. Boron vacancies are generated/activated using C ion irradiation with an energy of 10 keV and dose of $1\times 10^{14}~\mathrm{cm^{-2}}$. Further details can be found in Baber \textit{et al.}\cite{Baber_NL2022}.

\subsection{Experimental Setup}
Photoluminescence is excited using a 532 nm diode-pumped solid-state laser, modulated by an acousto-optic modulator. The laser is coupled to a long working distance objective lens (N.A.=0.8) which focuses the light to a diffraction-limited spot  $\sim1~\mathrm{\mu m}$ in diameter. The luminescence is collected with the same objective via a 750 nm long pass filter, to a fiber coupled single photon avalanche diode (SPAD). The intensity is recorded using a time-correlated single photon counting module. The microwave waveforms are generated using an arbitrary waveform generator, amplified (30 dB amplification, maximum output power 30 dBm) and applied to the CPW on the sample. The optical and microwave excitation and photon collection are synchronised using a digital pattern generator.

%BibTeX users: After compilation, comment out the following two lines and paste in
% the generated .bbl file.

\bibliography{scibib}

\bibliographystyle{Science}

\section*{Acknowledgments}
This work was supported by the Engineering and Physical Sciences Research Council [Grant numbers EP/S001557/1, EP/L015331/1 and EP/T017813/1] and Partnership Resource Funding from the Quantum Computing and Simulation Hub [EP/T001062/1]. Ion implantation was performed by Keith Heasman and Julian Fletcher at the University of Surrey Ion Beam Centre. We thank Dr J. P. Hadden for useful discussions at an early stage of the project.

%\section*{Supplementary materials}
%Materials and Methods\\
%Supplementary Text\\
%Figs. S1 to S3\\
%Tables S1 to S4\\
%References \textit{(4-10)}

% For your review copy (i.e., the file you initially send in for
% evaluation), you can use the {figure} environment and the
% \includegraphics command to stream your figures into the text, placing
% all figures at the end.  For the final, revised manuscript for
% acceptance and production, however, PostScript or other graphics
% should not be streamed into your compliled file.  Instead, set
% captions as simple paragraphs (with a \noindent tag), setting them
% off from the rest of the text with a \clearpage as shown  below, and
% submit figures as separate files according to the Art Department's
% instructions.

\clearpage

\end{document}